\documentclass[a4paper]{article}
\usepackage{a4}
\usepackage{fancyhdr}
\usepackage{graphicx}

\begin{document}
\pagestyle{fancy}
\setlength{\parskip}{2.0ex}
\setlength{\parindent}{0pt}

\title{Identification of the Nitrogen Interstitial as Origin of the 3.1 eV
  Photoluminescence Band in Hexagonal Boron Nitride}

\maketitle

\author{Elham Khorasani, Thomas Frauenheim, B\'{a}lint Aradi* and Peter De\'{a}k*}

Elham Khorasani\\
Bremen Center for Computational Materials Science, University of Bremen,
P.O. Box 330440, D-28334 Bremen, Germany.

Prof. Dr. Thomas Frauenheim\\
Beijing Computational Science Research Center (CSRC), 100193 Beijing, China.\\
Shenzhen JL Computational Science and Applied Research Institute, 518110
Shenzhen, China.\\
Bremen Center for Computational Materials Science, University of Bremen,
P.O. Box 330440, D-28334 Bremen, Germany.

Dr. B{\'{a}}lint Aradi\\
Bremen Center for Computational Materials Science, University of Bremen,
P.O. Box 330440, D-28334 Bremen, Germany.\\
Email Address: aradi@uni-bremen.de

Prof. Dr. Peter De{\'{a}}k\\
Bremen Center for Computational Materials Science, University of Bremen,
P.O. Box 330440, D-28334 Bremen, Germany.\\
Email Address: peter.deak@bccms.uni-bremen.de


\begin{abstract}
  
  Nitrogen interstitials (N$_\textsubscript{i}$) have the lowest formation
  energy among intrinsic defects of hexagonal boron nitride (hBN) under n-type
  and N-rich conditions. Using an optimized hybrid functional, which reproduces
  the gap and satisfies the generalized Koopmans condition, an
  N$_\textsubscript{i}$ configuration is found which is lower in energy than the
  ones reported so far. The (0/-) charge transition level is also much deeper,
  so N$_\textsubscript{i}$ acts as a very efficient compensating center in
  n-type samples. Its calculated photoluminescence (PL) at 3.0~eV agrees well
  with the position of an N-sensitive band measured at 3.1~eV. It has been also
  found that the nitrogen vacancy (V$_\textsubscript{N}$) cannot be the origin
  of the three boron electron (TBC) electron paramagnetic resonance (EPR) center
  and in thermal equilibrium it cannot even exist in n-type samples.

\end{abstract}

\section{Introduction}

Materials with well-defined defect centers are highly promising for next
generation devices, with applications ranging from electronics and photonics to
quantum computing \cite{ivady2020ab,kim2018photonic}. Hexagonal boron nitride
(hBN) is a wide-band-gap semiconductor that exhibits a rich number of band gap
levels which profoundly impact its electronic and optical properties. These can
act as recombination centers and emit light within a specific energy range. In
recent years, numerous color centers have been reported in hBN
\cite{tran2016robust, wong2015characterization}. However, the physical origin of
some of these centers is not yet fully understood
\cite{museur2008defect,andrei1976point}. First principle calculations can play a
crucial role in identifying the defects and providing accurate information about
their properties.

The luminescence properties of hBN have been studied mainly in recent years
\cite{museur2008defect,museur2009photoluminescence,du2015origin,silly2007luminescence,vuong2016phonon}. Museur
et al.\ reported three photoluminescence (PL) bands at the energies of 5.3~eV,
3.75~eV, and 3.1~eV for pyrolytic hBN (pBN) samples
\cite{museur2009photoluminescence}. pBN samples are high purity samples, free of
carbon and oxygen. The emission bands around 5.3~eV and 3.57~eV were found to be
donor-acceptor pair (DAP) type transitions. The nature of the PL band
observed at 3.1~eV in pBN samples was not elucidated. Du et al.\
\cite{du2015origin} investigated the PL spectra for a set of hBN epilayers grown
by metal organic chemical vapor deposition (MOCVD) under different ammonia
($\mathrm{NH}_3$) flow rates from 0.2 to 1.5 standard liters per minute
(SLM). They observed four emission peaks at energies of 4.12~eV, 3.92~eV,
3.72~eV, and 5.37~eV for hBN under the $\mathrm{NH}_3$ flow rate of 0.2 SLM. The
first three peaks were interpreted as a zero phonon line at 4.12~eV with two
phonon replicas. They observed a decrease in the emission intensity for the peak
at 4.12~eV upon increasing the $\mathrm{NH}_3$ flow rate while the change of the
peak at 5.37~eV was found to be small. They also observed the appearance of the
peak at 3.1~eV as the flow rate of the $\mathrm{NH}_3$ was increased to 1.5 SLM.

Besides PL spectroscopy, electron paramagnetic resonance (EPR) is a powerful
technique often used for identifying and studying defects in semiconductors and
insulators \cite{WATKINS}. Two types of paramagnetic centers were identified in
hBN
\cite{andrei1976point,geist1964paramagnetische,katzir1975point,moore1972electron}.
In one type, an unpaired electron interacts with three equivalent boron nuclei,
producing a 10-line EPR spectrum. This type is referred to as the three boron
center (TBC). In another type, an unpaired electron interacts with a single
boron nucleus, which gives rise to a 4-line EPR spectrum. This is referred to as
the one boron center (OBC). Based on experimental observation, the carbon
substitutional ($\mathrm{C}_\mathrm{N}$) and the nitrogen vacancy
($\mathrm{V}_\mathrm{N}$) were proposed
\cite{andrei1976point,moore1972electron,sajid2018defect} to model the TBC. In
the former, an unpaired electron was trapped at the carbon atom. This was
referred to as a carbon-associated TBC and was confirmed theoretically
\cite{sajid2018defect}. In the $\mathrm{V}_\mathrm{N}$ model, proposed by
experiments \cite{andrei1976point}, the unpaired electron is assumed to be
trapped in the nitrogen vacancy. This was referred to as an electron-irradiation
produced TBC. However, its nature is still controversial.

We have applied an optimized hybrid functional to calculate defects in hBN. Our
calculations show that the observed PL at 3.1~eV in bulk hBN is due to the
recombination of an electron with a hole trapped at a nitrogen interstitial,
$\mathrm{N}_\mathrm{i}$. Additionally, we have found that
$\mathrm{V}_\mathrm{N}$ cannot be the origin of the TBC center. In fact, in
n-type samples in equilibrium, $\mathrm{V}_\mathrm{N}$ turns out to be less
stable than $\mathrm{N}_\mathrm{i}$ even under extreme N-poor conditions,
indicating that $\mathrm{V}_\mathrm{N}$ can be created only by irradiation.

\section{Computational method}
\label{computational-method}

It was shown that the screened hybrid functional of Heyd, Scuseria, and
Ernzerhof (HSE) \cite{heyd2003hybrid,doi:10.1063/1.2204597}, with parameters
tuned to reproduce the relative position of the quasi-particle band edges and to
satisfy the generalized Koopman’s theorem (gKT), is capable of providing
the formation energy, the charge transition levels and the hyperfine interaction of
defects very accurately in traditional bulk semiconductors
\cite{deak2019optimized,deak2017choosing,han2017defect}.  In our previous work,
we have shown that such a functional can be optimized for layered compounds as
well \cite{deak2019defect} and HSE($\alpha,\mu$) with a mixing parameter
$\alpha$ = 0.3 and a screening parameter $\mu$ = 0.4 was found as the optimal
functional for bulk hBN.

Our calculations in the present work have been carried out with the Vienna Ab
initio Simulation Package, VASP 5.4.4, using the projector augmented wave method
\cite{kresse1994ab,kresse1996efficient,kresse1999ultrasoft}. A 420 (840) eV
cutoff was applied for the expansion of the wave functions (charge density) in
hBN. The convergence condition of 10$^{-4}$~eV was applied for the
self-consistent electronic energy. Van der Waals interactions between the layers
were taken into account by the Tkatchenko-Scheffler method
\cite{PhysRevLett.102.073005}, using $s_\mathrm{R}$ = 0.96.

The equilibrium geometry was determined for the primitive unit cell with a
$\Gamma$-centered $6 \times 6\times 6$ Monkhorst-Pack (MP) $k$-point set
\cite{monkhorst1976special}, based on constant volume relaxations and fitting to
Murnaghan's equation of state \cite{Murnaghan244}. We describe bulk hBN using an
orthogonal supercell of 120 atoms (5\textbf{a}$_{1}$, 3\textbf{a}$_{1}$ +
6\textbf{a}$_{2}$, \textbf{a}$_{3}$) at the calculated lattice constants (see
our previous work \cite{deak2019defect}), using the $\Gamma$-point
approximation. (Here $\textbf{a}_{1}$, $\textbf{a}_{2}$, and $\textbf{a}_{3}$
are the primitive unit vectors). The geometries of the defects in the supercell
were relaxed at fixed lattice constants, using a force criterion of 0.01~eV/\AA.

We have calculated the formation energy and the transition levels of the
intrinsic defects $\mathrm{N}_\mathrm{i}$ and $\mathrm{V}_\mathrm{N}$. The
chemical potentials $\mu_{\mathrm{N}}$ and $\mu_{\mathrm{B}}$ are related in
equilibrium, as $\mu_{\mathrm{BN}}=\mu_{\mathrm{B}}+\mu_{\mathrm{N}}$ where
$\mu_\mathrm{BN}$ is the energy of hBN per formula unit. For N-rich conditions
the chemical potential of nitrogen was taken as the energy of the nitrogen atom
in the molecule (at a temperature of 300~K and a pressure of 10$^5$~Pa). For the
B-rich case $\mu_\mathrm{B}$ corresponds to the energy of the B atom in the
solid phase. For charged defects, total energies were corrected \textit{a
  posteriori}, using the SLABCC code \cite{tabriz2019slabcc,SLABCC} to eliminate
the artificial interaction between the repeated charges. SLABCC is based on the
scheme proposed by Komsa and Pasquarello~\cite{PhysRevX.4.031044}. For the
correction, the static dielectric constant $\varepsilon_{0}$ should be used in
principle if the ionic cores are allowed to relax. However, it was shown that,
due to the explicit screening in the supercell, the high-frequency value
$\varepsilon^{\infty}$ gives a better approximation \cite{PhysRevB.95.075208},
therefore $\varepsilon^{\infty}_{\perp} = 4.95$ and
$\varepsilon^{\infty}_{\parallel} = 4.10$ \cite{geick1966normal} were used.

To explain the observed PL in bulk hBN, we have investigated electron
recombination with a hole trapped at $\mathrm{N}_\mathrm{i}$. Within the
accuracy of the calculation (0.1~eV), a shallow donor state cannot be
distinguished energetically from the bottom of the conduction band. Therefore,
we simulate a donor-acceptor recombination from a shallow donor to a deep
acceptor by the recombination of a bound exciton, i.e., from the conduction band
edge to the acceptor level \cite{du2015origin}. To calculate the PL, we first
relaxed the equilibrium geometry of the bound exciton under the constraint of
the orbital occupations, i.e., with one hole at the defect level and with one
electron in the conduction band minimum (CBM).  To consider the delocalization
of the electron, the total energy of the initial state was recalculated with a
(non-$\Gamma$-centered) $2 \times 2 \times 2$ Monkhorst-Pack set at the relaxed
geometry of the bound exciton in the $\Gamma$ approximation. The energy of the
final state after recombination was calculated at fixed geometry, using the same
$2 \times 2 \times 2$ $k$-point set. The PL energy is the difference between the
two total energies.

Hyperfine interaction was calculated for $\mathrm{V}_\mathrm{N}$. Boron has
two isotopes, $^{10}{\mathrm{B}}$ and $^{11}{\mathrm{B}}$, with the natural
abundances of $18.83\%$ and $81.17\%$, respectively. These isotopes have
different nuclear spins and magnetic moments. In this work, the hyperfine
splitting was compared to the experimental data measured in
$^{11}{\mathrm{B}}$-enriched samples. The presented results were obtained for
the isotope $^{11}{\mathrm{B}}$ with $I=3/2$. The nuclear gyromagnetic ratios of
13.66 and 3.08 were used for boron and nitrogen, respectively
\cite{NUCLEARRATIO}.

\section{Results and Discussion}

\subsection{Nitrogen interstitial}

\begin{figure}[htb]
  \centering
  \includegraphics*[width=0.5\linewidth]{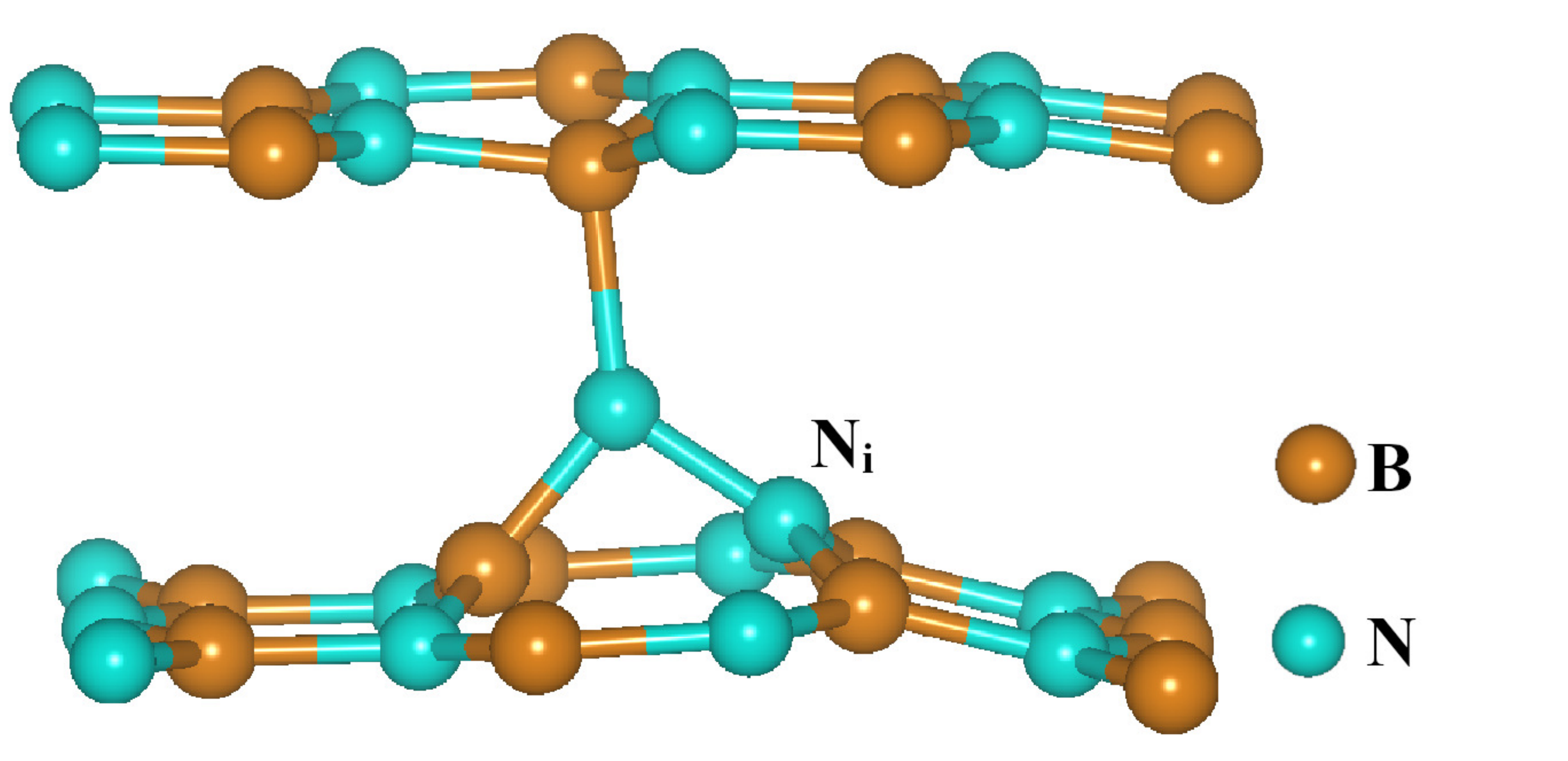}
  \caption{Atomic structure of the intrinsic defect $\mathrm{N}_\mathrm{i}$ in
    the lowest energy configuration. The $\mathrm{N}_\mathrm{i}$-$\mathrm{N}$
    bondlength, 1.38~\AA\ corresponds to a single bond. $\mathrm{N}_\mathrm{i}$
    also binds to a boron atom of the top layer.}
  \label{fig:Fig1}
\end{figure}

In earlier works, the nitrogen interstitial was predicted to be in a [001] split
interstitial configuration with a lattice N atom
\cite{PhysRevB.97.214104,wang2014hybrid}, and was found to be the intrinsic
defect with the lowest formation energy under N-rich conditions in n-type
material. We have found, however, that the configuration shown in
\textbf{Figure~\ref{fig:Fig1}} is lower in energy by 0.70~eV. The axis of the
split interstitial tilts away from [001], so both nitrogen atoms are ideally
threefold coordinated. The $\mathrm{N}_\mathrm{i}$-$\mathrm{N}$ bondlength,
1.38~\AA\ corresponds to a single bond, and $\mathrm{N}_\mathrm{i}$ also binds
to a boron atom of the top layer. Simple electron counting shows that the
$p$-orbital of $\mathrm{N}_\mathrm{i}$, which is orthogonal to the plane of its
three bonds, contains only one electron.

\begin{figure}
  \centering
  \includegraphics[width=0.45\columnwidth]{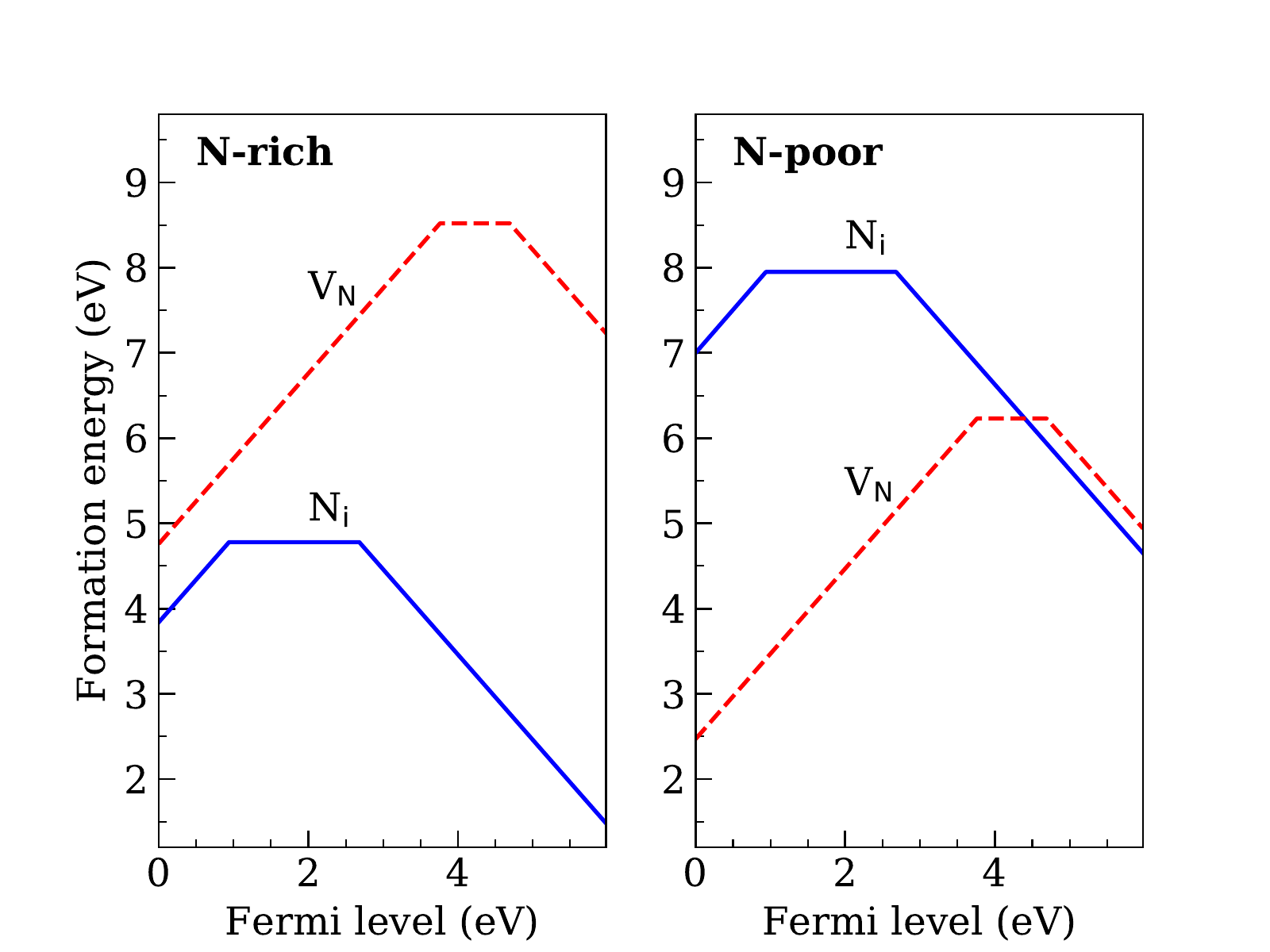}
  \caption{Formation energies of the intrinsic point defects
    $\mathrm{N}_\mathrm{i}$ and $\mathrm{V}_\mathrm{N}$ in bulk hBN as a
    function of the Fermi level under (left) N-rich and (right) N-poor
    conditions. Solid blue and dashed red lines represent the defects
    $\mathrm{N}_\mathrm{i}$ and $\mathrm{V}_\mathrm{N}$, respectively. The
    energy of the valence band maximum has been set to zero.}
  \label{fig:Fig2}
\end{figure}

\textbf{Figure~\ref{fig:Fig2}} shows the formation energy calculated for the
intrinsic defects $\mathrm{N}_\mathrm{i}$ and $\mathrm{V}_\mathrm{N}$ in
different charge states, as a function of the Fermi level position between the
valence band maximum (VBM) and the conduction band minimum (CBM). The adiabatic
charge transition levels for $\mathrm{N}_\mathrm{i}$ are obtained at
$\mathrm{E}(+\slash0)=$ 0.94~eV and $\mathrm{E}(0\slash-)=$ 2.68~eV with respect
to the valence band edge. The latter is much deeper than the value obtained in a
previous work \cite{PhysRevB.97.214104}. Weston et al.\ predicted
$\mathrm{N}_\mathrm{i}$ to be the major compensating center in n-type samples
\cite{PhysRevB.97.214104}. Our finding confirms that and the lower formation
energy indicates an even stronger compensating effect.

The PL energy was calculated for the configuration shown in
Figure~\ref{fig:Fig1} as described in Section
\ref{computational-method}. Assuming the donor level to be 100~meV under the
CBM, our calculated PL energy is at 3.0~eV. This can explain the origin of the
N-related observed peak at 3.1 eV in hBN in
Ref. \cite{du2015origin,museur2009photoluminescence}.

\subsection{Nitrogen Vacancy}

\begin{figure}
  \includegraphics[width=\linewidth]{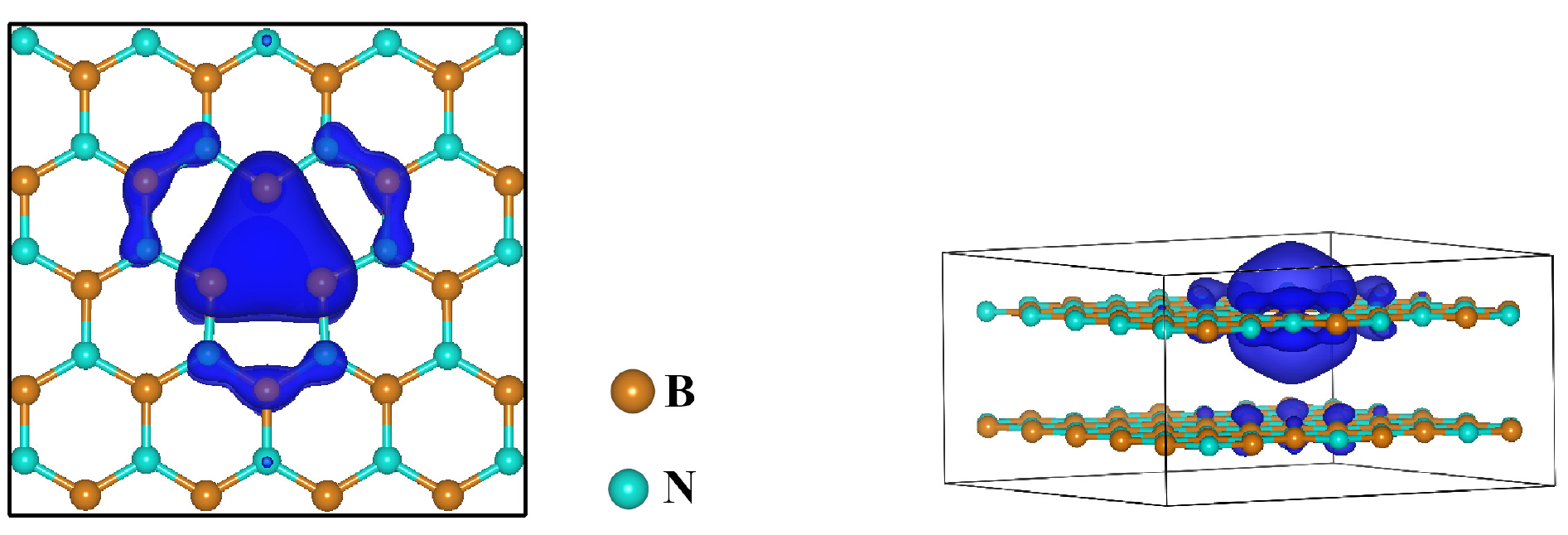}
  \caption{The spin density calculated for the $\mathrm{V}_\mathrm{N}$ defect as
    model for the TBC in bulk hBN (isovalue: 0.002 1/{\AA}$^{3}$) in (left) top
    view and (right) side view. The spin density concentrated on the three
    neighbouring boron atoms produces the calculated hyperfine couplings.}
  \label{fig:Fig3}
\end{figure}

The similarities between optical emitters in hBN and the nitrogen-vacancy
($\mathrm{NV}$) center in diamond, have brought a lot of attention to vacancy
defects in hBN. Earlier experimental studies proposed $\mathrm{V}_\mathrm{N}$ as
the source of the TBC \cite{andrei1976point}, which was produced by electron
irradiation. $\mathrm{V}_\mathrm{N}$ introduces a gap level with an unpaired
electron. Our calculations show that the electron is on $p$-orbitals orthogonal
to the lattice planes, with a symmetrical charge distribution on the three
neighboring boron atoms. \textbf{Figure~\ref{fig:Fig3}} shows the spin
distribution. The calculated hyperfine coupling constants $A$ are given in Table
\ref{tab1}. The calculated average principle value of the hyperfine interaction
tensor is obtained at 34.23 MHz, which is significantly smaller than the value
of 117.06 MHz measured for the TBC in carbon-free samples after irradiation. It
has been shown that core spin polarization may have a significant effect on the
hyperfine interaction calculated, providing accurate results in various
semiconductors \cite{PhysRevB.88.075202}. Taking the core contribution into
account, the calculated magnitude of 34.23~MHz, which only involved the valence
electronic spin density, reduced to 16.09~MHz, demonstrating that the
$\mathrm{V}_\mathrm{N}$ can not be the origin of the three boron center.

The nitrogen instersitial defect, $\mathrm{N}_\mathrm{i}$ is expected to be more
stable than $\mathrm{V}_\mathrm{N}$ in N-rich samples. This is also confirmed by
Figure~\ref{fig:Fig2}. However, as demonstrated by the same figure, even under
extreme N-poor conditions, $\mathrm{V}_\mathrm{N}$ can only exist in p-type
samples in thermal equilibrium. Actually, there is no hard experimental evidence
for the existence of $\mathrm{V}_\mathrm{N}$. In the earlier PL studies
\cite{museur2008defect,du2015origin,silly2007luminescence,watanabe2004direct},
the nitrogen vacancy was suggested as a possible source of the 4.1 eV emission
band in bulk hBN. Later on, however, a theoretical study showed that the carbon
dimer defect $\mathrm{C}_\mathrm{B}\mathrm{C}_\mathrm{N}$ is responsible for the
4.1 eV emission in hBN \cite{mackoit2019carbon}. Jin et
al. \cite{jin2009fabrication} used ultrahigh-resolution transmission electron
microscope imaging. They created defects by an electron beam irradiation and
found that the created defects are of the boron vacancy
($\mathrm{V}_\mathrm{B}$) type.

\begin{table}
  \centering
  \caption{Hyperfine coupling constants for the model defect
    $\mathrm{V}_\mathrm{N}$. The average value of $A_{\mathrm{ave}}$ was
    calculated as ($A_{\mathrm{xx}}+ A_{\mathrm{yy}}+ A_{\mathrm{zz}})/3$ in
    MHz.}
  \begin{tabular}[htbp]{@{}lllll@{}}
    \hline
    $A_{\mathrm{xx}}$ & $A_{\mathrm{yy}}$ & $A_\mathrm{zz}$ & $A_\mathrm{ave}$ &
    $A_{\mathrm{ave}}$(experimental)\footnote{andrei1976point}\\ \\
    \hline
    21.21  & 20.81  & 60.65 & 34.23 & 117.06  \\
    \hline
  \end{tabular}
  \label{tab1}
\end{table}

\section{Conclusion}

We have calculated the intrinsic defects, nitrogen interstitial and nitrogen
vacancy, using the optimal HSE (0.4, 0.3) functional for hBN as obtained in our
previous work. We have found an $\mathrm{N}_\mathrm{i}$ configuration which is
considerably lower in energy than the ones reported so far. The calculated
formation energies show that $\mathrm{N}_\mathrm{i}$ is the intrinsic defect
with the lowest formation energy under N-rich conditions. Its
acceptor level is deep and behaves as a compensating center. The calculated PL
energy of 3.0~eV for $\mathrm{N}_\mathrm{i}$ is in good agreement with the
experimentally observed N-related band at 3.1 eV
\cite{museur2009photoluminescence,du2015origin}. We have investigated the
nitrogen vacancy as well. Our result indicates that $\mathrm{V}_\mathrm{N}$ is
less stable than $\mathrm{N}_\mathrm{i}$ in n-type samples even under extreme
N-poor conditions and should be observable only in irradiated samples.
The hyperfine interactions have been calculated for $\mathrm{V}_\mathrm{N}$,
which was suspected to be a source of the three boron centers produced under an
electron-irradiation in carbon-free samples. Our results indicate that
$\mathrm{V}_\mathrm{N}$ cannot be the origin of the three boron center in bulk
hBN.

\medskip

\textbf{Acknowledgements} \par
This work was supported by the Deutsche Forschungsgemeinschaft (DFG) within the
research training group (RTG) 2247.

\end{document}